\documentclass[prl,aps,amsmath,amssymb,showpacs,superscriptaddress,twocolumn]{revtex4}

\usepackage{graphics}
\usepackage{dcolumn}
\usepackage{bm}
\usepackage[T1]{fontenc} 

\begin{document}

\title{Zeeman splitting in ballistic hole quantum wires}

\author{R. Danneau}\email[Corresponding author.\\Electronic address:
r.danneau@unsw.edu.au]{}
\author{O. Klochan}
\author{W.R. Clarke}
\author{L.H. Ho}
\author{A.P. Micolich}
\author{M.Y. Simmons}
\author{A.R. Hamilton}
\affiliation{School of Physics,University of New South Wales,
Sydney 2052, Australia.}
\author{M. Pepper}
\author{D.A. Ritchie}
\affiliation{Cavendish Laboratory, Madingley Road, Cambridge, CB3
OHE, United Kingdom.}
\author{U. Z\"{u}licke}
\affiliation{Institute of Fundamental Sciences and MacDiarmid
Institute for Advanced Materials and Nanotechnology, Massey
University, Palmerston North, New Zealand.}


\begin{abstract}

We have studied the Zeeman splitting in ballistic hole quantum
wires formed in a (311)A quantum well by surface gate confinement.
Transport measurements clearly show lifting of the spin degeneracy
and crossings of the subbands when an in-plane magnetic field $B$
is applied parallel to the wire. When $B$ is oriented
perpendicular to the wire, no spin-splitting is discernible up to
$B$ = 8.8 T. The observed large Zeeman splitting anisotropy in our
hole quantum wires demonstrates the importance of
quantum-confinement for spin-splitting in nanostructures with
strong spin-orbit coupling.

\end{abstract}


\maketitle

Studying the spin degree of freedom of charge carriers in
semiconductors has become an area of significant current interest,
not only for fundamental understanding of spin, but also for
potential applications that use spin, rather than charge, in
electronic components \cite{zutic2004}. Spin polarized currents
can be created by applying magnetic fields, using magnetic leads
or ferromagnetic semiconductors. Alternatively, the intrinsic
coupling between spin state and orbital motion of quantum
particles opens up intriguing possibilities for implementing a
spin-electronics paradigm. For example, it has been proposed to
tune the spin splitting by using an external electric-field in
systems exhibiting strong spin-orbit coupling \cite{byshkov1984}.
In such devices, a spin polarized current could be manipulated in
a ballistic channel by only tuning a gate voltage
\cite{datta1990}. As valence-band states are predominantly p-like
(unlike conduction-band states which are s-like), spin-orbit
effects are particularly important in confined hole systems based
on GaAs \cite{winkler2003}. This makes holes in GaAs especially
interesting for studies of spin-controlled devices. In addition,
the fact that holes near the valence-band edge are spin 3/2
particles leads to intriguing quantum effects such as the
suppression of Zeeman splitting for in-plane field directions in
typical two-dimensional (2D) hole systems \cite{vankeresten1990}.
How further confinement of holes moving in a narrow wire affects
their peculiar spin properties has not been investigated in detail
before, which motivates our present study. Furthermore, Zeeman
splitting of one-dimensional (1D) subband bottoms can be {\em
directly\/} measured using the phenomenon of conductance
quantization
\cite{patel1991a,thomas1996,thomas1998,daneshvar1997}.

We have performed an experimental study of the Zeeman splitting of
quantum wires in the ballistic regime, formed by a lateral
confinement of a 2D heavy-hole (HH) system that was grown on the
(311)A surface of a GaAs/Al$_{\mathrm{x}}$Ga$_{1-\mathrm{x}}$As
heterostructure. Previous works on 2D systems
\cite{papadakis2000,winkler2000,winkler2003} identified an
anisotropic effective Land\'{e} \emph{g}-factor $g^{*}$ for
in-plane magnetic fields applied in the $[\overline{2}33]$ and
$[01\overline{1}]$ directions. In our 1D system which is aligned
with the $[\overline{2}33]$ direction, when we apply an in-plane
$B$ we measure a much larger $g^{*}$ anisotropy between
parallel-to-the-wire ($B_{\parallel}$) and
perpendicular-to-the-wire ($B_{\perp}$) than is predicted (and
observed) for 2D HH systems \cite{note0}. We attribute these
observations to the interplay between quantum confinement and
strong spin-orbit coupling present in the valence band. Our
results show that it is possible to engineer magnetic (Zeeman)
splitting by tuning the electric confinement in hole
nanostructures.

\begin{figure*}[htbp]
\scalebox{0.5} {\includegraphics{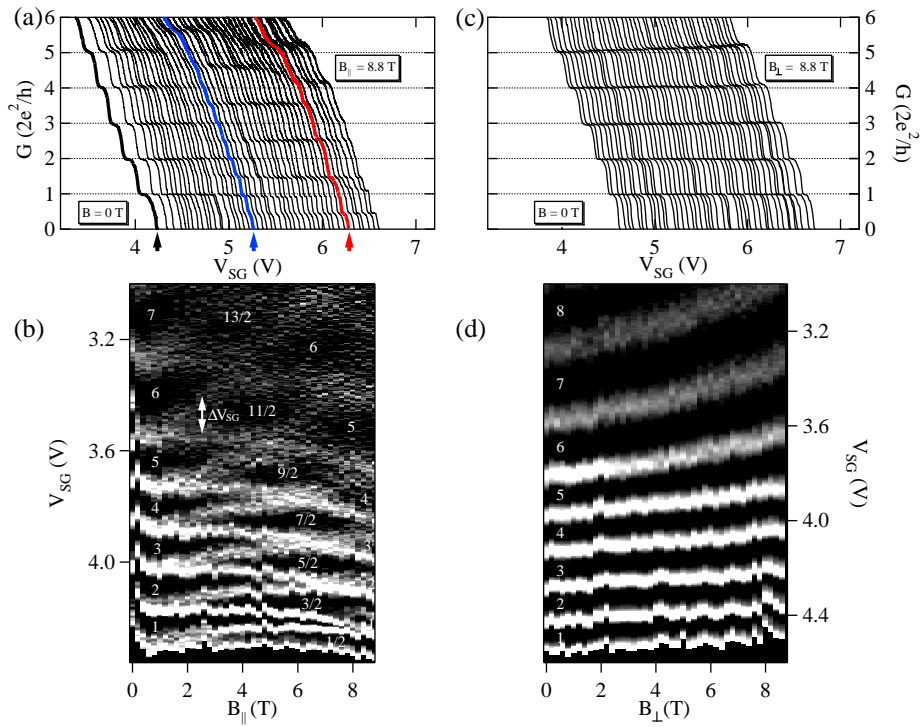}}{\caption{(a)
Differential conductance $G$, corrected for a series resistance,
of the quantum wire for different in-plane magnetic fields
parallel to the wire ($B_{\parallel}$) from 0 T to 8.8 T in steps
of 0.2 T (from left to right). \emph{T} = 20 mK, back and middle
gates are at 2.5 V and  -0.5 V, respectively. All curves are
shifted by 0.05 V for clarity. The second thicker curve (middle
arrow) corresponds to $B_{\parallel} = B_{\mathrm{R}}$ when
subbands are completely spin resolved; the third thicker curve
(rightmost arrow) corresponds to $B_{\parallel} = B_{\mathrm{C}}$
when the subbands cross. (b) Corresponding transconductance
grayscale as a function of $V_{\mathrm{SG}}$; black regions
correspond to low transconductance (conductance plateaus, labelled
with $G$ in units of $2e^{2}/h$), white regions correspond to high
transconductance (subband edges). (c) $G$ of the same quantum wire
for different in-plane magnetic fields perpendicular to the wire
($B_{\perp}$) under similar experimental conditions. (d)
Corresponding transconductance grayscale as a function of
$V_{\mathrm{SG}}$ and $B_{\perp}$.}} \label{pinchoffbilayer4K}
\end{figure*}

In our experiments, we used the same 1D hole bilayer system as
described previously \cite{danneau2005} and we have measured the
differential conductance in the top wire of the bilayer
\cite{note1}. Measurements were done in a dilution refrigerator
with a base temperature of $T$ = 20 mK. Side gates and a middle
gate used to form the quantum wire are aligned along the
$[\overline{2}33]$ direction. Electrical measurements have been
performed using standard low-frequency ac lock-in techniques with
an excitation voltage of 20 $\mu$V at 17 Hz. We used two side
gates to create the 1D channels and the back and middle gates to
control the density and the confinement potential as described in
\cite{danneau2005}.

Figure 1 (a) shows the differential conductance, $G =
\mathrm{d}I/\mathrm{d}V$ as a function of side gate voltage
$V_{\mathrm{SG}}$ for different in-plane magnetic fields parallel
to the wire $B_{\parallel}$. Clean conductance quantization is
measured \cite{note2} and, with respect to $B_{\parallel}$ ,
Zeeman splitting is clearly seen. We observe the progressive
evolution of the 1D subbands from the spin degenerate steps in
units of $2e^2/h$ at $B_{\parallel} = 0$ T (leftmost arrow), to
the complete spin-resolved quantized steps in units of $e^2/h$ at
$B_{\parallel} = B_{\mathrm{R}} \approx 3.6$ T (middle arrow).
Further increasing $B$ causes the 1D non-degenerate subbands to
cross, leading to conductance plateaus quantized in units of
half-odd multiple values of $2e^2/h$ at $B_{\parallel}=
B_{\mathrm{C}} \approx 7.6$ T (rightmost arrow). In Fig.\ 1 (b),
we plot the transconductance
$\mathrm{d}G/\mathrm{d}V_{\mathrm{SG}}$ as a function of side gate
voltage and $B$ (the derivative has been numerically calculated
from $G$ corrected for a series resistance). Crossings between
non-degenerate 1D subband edges are clearly visible (white regions
of the grayscale) as well as the diamond shape parts (in black)
representing the conductance plateaus.

After thermal cycling and sample re-orientation, we have measured
the differential conductance for an in-plane magnetic fields
perpendicular to the wire $B_{\perp}$. It is surprising to see in
Figs.\ 1(c) and 1(d) that the well-quantized conductance steps are
not affected by $B_{\perp}$ up to 8.8 T. The transconductance
grayscale shown in Fig.\ 1 (d) highlights that no Zeeman splitting
is seen when the magnetic field is aligned perpendicular to the
wire.

In Fig.\ 2 (a), we present a schematic view of the effect of an
in-plane magnetic field on the spin degenerate 1D subbands. We
assume that the splitting of a 1D energy subband is linear in
magnetic field \cite{fang1968} and follows the equation $\Delta
E_{\mathrm{N}}=g^{*}_{\mathrm{N}}\mu_{\mathrm{B}}B$, where $\Delta
E_{\mathrm{N}}$ is the Zeeman energy splitting of the
N$^{\mathrm{th}}$degenerate subband, $g^{*}_{\mathrm{N}}$ is the
effective Land\'{e} \emph{g}-factor of the
N$^{\mathrm{th}}$degenerate subband, $\mu_{\mathrm{B}}$ is the
Bohr magneton and $B$ is the applied magnetic field. In this
diagram, the lines represent the subband edges as measured in
Fig.\ 1 (b). However, Zeeman splitting measurements, \emph{i.e.}
splitting of the transconductance peaks, is given in units of
$\Delta V_{\mathrm{SG}}(B)$. It is possible to convert $\Delta
V_{\mathrm{SG}}(B)$ to $\Delta E_{\mathrm{N}}(B)$ and extract
values of $g^{*}$, by combining Zeeman splitting measurements and
source-drain bias $V_{\mathrm{SD}}$ spectroscopy \cite{patel1991}.
In Fig.\ 2 (b), we present a schematic of the transconductance as
a function of the source-drain bias $V_{\mathrm{SD}}$ that allows
the extraction of the subband spacings by tuning chemical
potentials of the source and drain with respect to the 1D subbands
($\Delta E_{\mathrm{N,N+1}}=eV_{\mathrm{SD}}$ at subband
crossing), according to
\cite{patel1991,glazman1989,martinmoreno1992}. Fig.\ 3 corresponds
to a typical transconductance grayscale as a function of
$V_{\mathrm{SD}}$: clear half plateaus
\cite{patel1991,glazman1989,martinmoreno1992} (\emph{i.e.}
conductance plateau are quantized in units of half-odd multiple of
$2e^2/h$) are seen for high values of $V_{\mathrm{SD}}$ and 1D
subband edges (in white) are visible.

\begin{figure}[htbp]
\scalebox{0.4}{\includegraphics{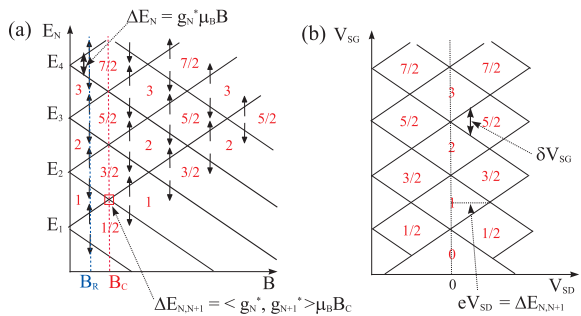}} \caption{(a)
Schematic diagram of the Zeeman effect in a 1D system due to an
in-plane magnetic field $B$: solid lines correspond to the 1D
subband edges, spin orientations are given by the arrows. The
first dotted line corresponds to $B$ at fully resolved spin
splitting $B_{\mathrm{R}}$: at this stage, conductance steps are
quantized in units of $e^2/h$. The second dotted line corresponds
to 1D subband crossings at $B_{\mathrm{C}}$: at this stage,
conductance steps are quantized of half-odd multiples of $2e^2/h$.
(b) Schematic of the transconductance for different
$V_{\mathrm{SD}}$: with this technique, it is possible to extract
the spacing between two consecutive subbands (see
\cite{glazman1989,patel1991,martinmoreno1992}). Combined with
Zeeman splitting measurement, $g^{*}$ can then be found.}
\label{pinchoffbilayer4K}
\end{figure}

A common way to extract $g^{*}$ is to compare the crossing of two
subband edges due to Zeeman splitting with that obtained from
source-drain bias techniques \cite{patel1991a} (if these crossings
appear for the same $V_{\mathrm{SG}}$). We can relate bias voltage
and energy: $\Delta E_{\mathrm{N,N+1}}= \langle
g^{*}_{\mathrm{N}},g^{*}_{\mathrm{N+1}} \rangle
\mu_{\mathrm{B}}B_{\mathrm{C}}= eV_{\mathrm{SD}}$ where $\langle
g^{*}_{\mathrm{N}},g^{*}_{\mathrm{N}+1}\rangle$ is the average
value of the effective Land\'{e} \emph{g}-factor for degenerate
subbands N and N+1 and $B_{\mathrm{C}}$ is the magnetic field at
subband crossing (see Fig.\ 2 (a) and (b) and $\langle
g^{*}_{\mathrm{N}},g^{*}_{\mathrm{N}+1}\rangle$ values
corresponding to solid circles in Fig.\ 4 (b)). Another way to
relate gate voltage and energy is to combine the splittings of
transconductance grayscale lines (\emph{i.e.} transconductance
peaks) from both source-drain bias and Zeeman effect (\emph{i.e.}
combine respectively $\delta V_{\mathrm{SG}}(V_{\mathrm{SD}})$ and
$\Delta V_{\mathrm{SG}}(B)$). We can use the basic relation for
the linear Zeeman splitting of a 1D subband $\partial \Delta
E_{\mathrm{N}}/\partial B=\partial \Delta E_{\mathrm{N}}/\partial
V_{\mathrm{SG}}\times
\partial V_{\mathrm{SG}}/\partial
B =g^{*}_{\mathrm{N}}\mu_{\mathrm{B}}$. Indeed, Zeeman splitting
of the transconductance lines (\emph{i.e.} transconductance peaks)
give the 1D degenerate subband splitting differential change with
respect to $B$, $\Delta V_{\mathrm{SG}}(B)$ (see Fig.\ 2 (a)). The
slope of $\Delta V_{\mathrm{SG}}(B)$ corresponds to $\partial
V_{\mathrm{SG}}/\partial B$. $V_{\mathrm{SD}}$ spectroscopy
provides the conversion factor between gate voltage and energy,
given by the slope of $\delta V_{\mathrm{SG}}(V_{\mathrm{SD}})$,
$\delta V_{\mathrm{SG}}/eV_{\mathrm{SD}}$. Finally, $g^{*}$ for
the N$^{th}$ subband is
$g^{*}_{\mathrm{N}}\mu_{\mathrm{B}}=eV_{\mathrm{SD}}/\delta
V_{\mathrm{SG}}\times\Delta V_{\mathrm{SG}}/B$ ($g^{*}$ values
extracted from this second technique correspond to the open
circles in Fig.\ 4 (b)).

\begin{figure}[htbp]
\scalebox{0.4}{\includegraphics{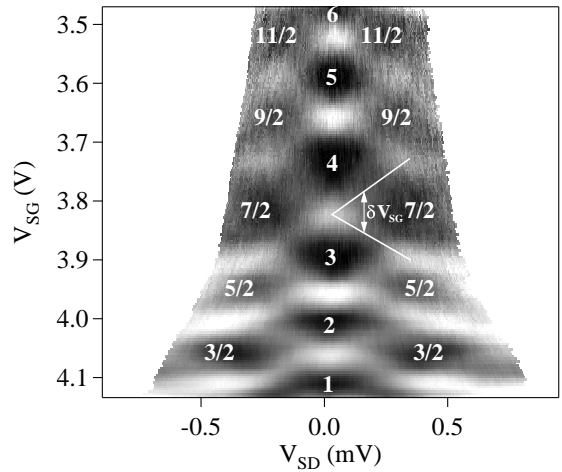}} \caption{Nonlinear
transconductance grayscale at \emph{T} = 20 mK, back and middle
gates are at 2.5 V and  -0.5 V, respectively, and \emph{B} = 0 T
as a function of $V_{\mathrm{SD}}$ \cite{note3}. The black parts
correspond to low transconductance (plateaus). Quantized plateaus
in units of $2e^2/h$ at zero $V_{\mathrm{SD}}$ and extra plateaus
for half-odd multiple values of $2e^2/h$ at non-zero
$V_{\mathrm{SD}}$ are labelled. Crossings of 1D subband edges are
the white parts.} \label{figure4APL}
\end{figure}

In Fig.\ 4 (a) we show the splitting of the transconductance
peaks, \emph{i.e.} the splitting of the 1D subband edges,
represented by $\Delta V_{\mathrm{SG}}$ as a function of $B$ for
the first five 1D subbands. The spin-splitting can be considered
to be linear in $B$ for the four first subbands. The fifth 1D
subband clearly shows a deviation from a purely linear Zeeman
splitting as the 1D system is becoming more 2D. Fig.\ 4 (b)
displays $g^{*}$ as a function of the degenerate 1D subbands index
N: Remarkably, values and behavior of $g_{\parallel}^{*}$(N) for
$B_{\parallel}$ parallel to the wire are similar as found in
previous results for electron quantum wires
\cite{patel1991a,thomas1996,thomas1998}. Because no sign of the
beginning of spin-splitting is detected for $B$ applied
perpendicular to the wire up to $B_{\perp} = 8.8$ T, we only can
provide maximum values of $g_{\perp}^{*}$ in this orientation; the
minimum $g^{*}$ ratio $g^{*}_{\parallel}$/$g^{*}_{\perp}$ can be
estimated \emph{at least} to be 4.5. This anisotropy is
significantly stronger than in 2D HH systems
\cite{winkler2003,papadakis2000,winkler2000}. Here, we provide an
explanation with detailed calculations to be presented elsewhere
\cite{zulicke2006}.

\begin{figure}[htbp]
\scalebox{0.325}{\includegraphics{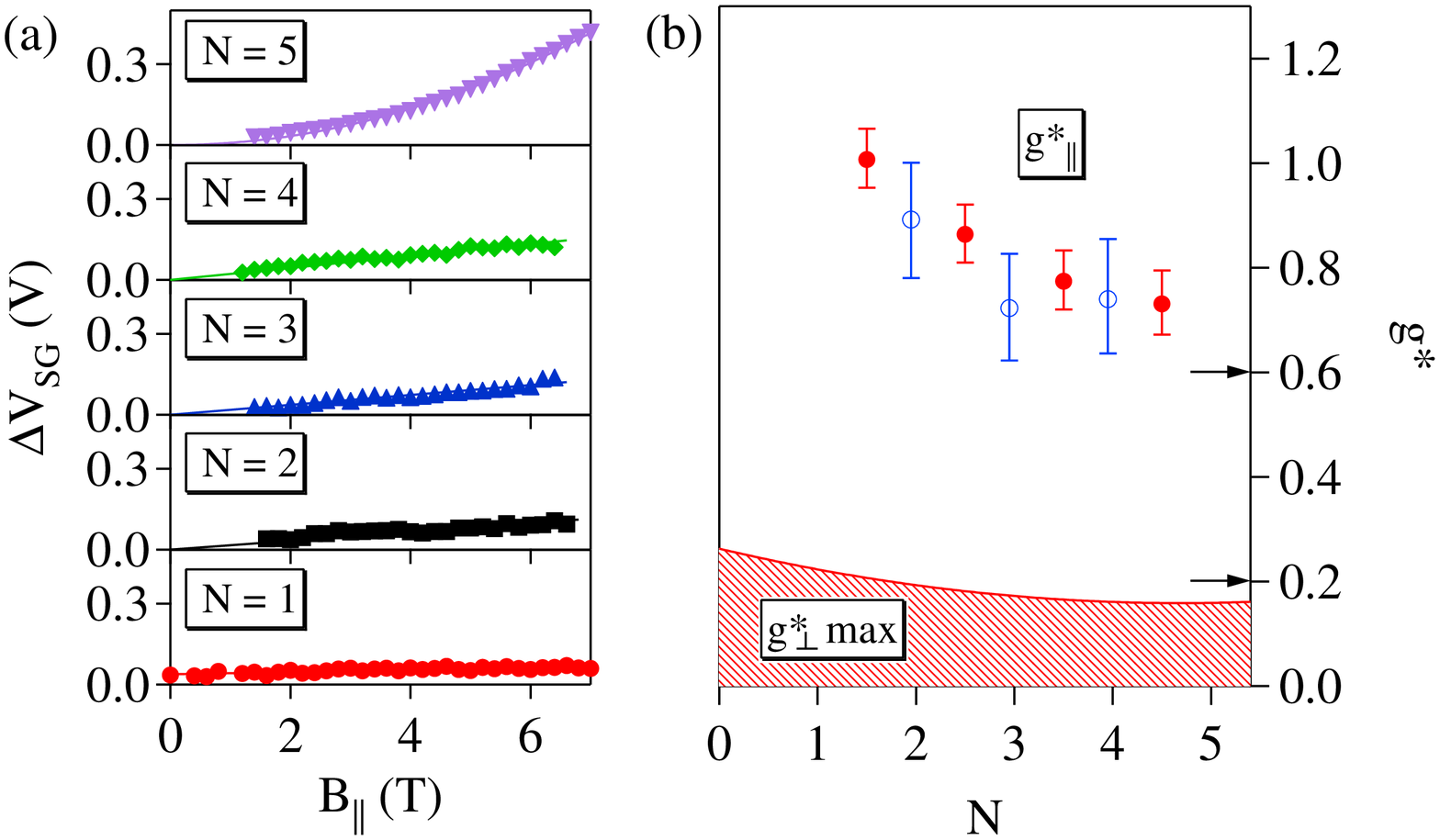}} \caption{(a)
$\Delta V_{\mathrm{SG}}$ \emph{vs} $B_{\parallel}$ for the five
first subbands; (b) g* as a function of the subband index N
extracted from the Zeeman splitting measurement and the
$V_{\mathrm{SD}}$ spectroscopy: Solid circles are the average
effective Land\'{e} \emph{g}-factor in the direction parallel to
the wire, between two consecutive subbands $\langle
g^{*}_{\mathrm{N}},g^{*}_{\mathrm{N}+1}\rangle$ given by
subband-edge crossing measurements \cite{patel1991a}. Open circles
are extracted from the slope of the subband edge splitting
differential change technique for the $B_{\parallel}$ pointing
along the wire. The upper line of the striped part represents an
upper bound of $g^{*}$ for $B_{\perp}$ perpendicular to the wire:
this limit is determined by comparing the starting point of the 1D
subband splittings ($B_{\parallel}\approx 1.9$ T) and the maximum
value of $B_{\perp}$ ($= 8.8$ T). The arrows define the absolute
values of $g^{*}$ calculated for a 20 nm quantum well of HH grown
on (311)A surface \cite{winkler2003,winkler2000}: the upper and
lower arrows mark $g^{*}$ for $B$ pointing along
$[\overline{2}33]$ and $[01\overline{1}]$, respectively.} \label{}
\end{figure}

In a quantum-confined structure, Zeeman splitting of HH states is
suppressed unless the magnetic field is applied parallel to the
natural quantization axis for total angular momentum $\hat{J}$
\cite{winkler2003}. \emph{This is a result of strong spin-orbit
coupling in the valence band}. For a 2D system, this axis is
perpendicular to the plane in which the holes are confined. As a
result, $g^{*}$ for the field direction perpendicular to the plane
is typically at least an order of magnitude larger than that for
magnetic fields applied in the plane. Also, the cubic crystal
anisotropy gives rise to an anisotropy of $g^{*}$ for orthogonal
in-plane field directions \cite{winkler2000}. In our 1D hole
system, in addition to the large anisotropy of $g^{*}$ for
$B_{\parallel}$ and $B_{\perp}$, we also measured a considerably
{\em larger\/} value of $g^{*}_{\parallel}$ for the lowest 1D
subbands than was predicted theoretically \cite{winkler2003} for
the 20 nm wide quantum-well as used in our sample. These findings
can be explained by the fact that a 1D confinement tends to favor
a quantization axis of {\em $\hat{J}$ parallel to the wire\/},
\emph{i.e.} perpendicular to the directions in which hole motion
is quantized. Hence a large Zeeman splitting results when $B$ is
applied in the same direction. For $B$ applied perpendicular to
the wire, no direct Zeeman coupling between HH states exists, and
their spin splitting arises only as a second-order effect from the
HH-LH couplings that are due to $B$ and confinement. It is
therefore suppressed by the confinement-induced energy splitting
between the HH and LH 1D subbands. For our hole quantum wires, the
confinement in the $[311]$ direction given by the quantum well can
still be expected to be stronger than the confinement in the
lateral direction, \emph{i.e.} $[01\overline{1}]$. Nevertheless,
the monotonic increase of $g^{*}_{\parallel}$ over the 2D value of
0.6 \cite{winkler2003,winkler2000}, which is exhibited in Fig.\ 4,
as the wire is made narrower (\emph{i.e.}, for smaller subband
index $N$) clearly indicates the expected trend. Note also that
the direct HH Zeeman splitting for fields parallel to the wire
could be enhanced by exchange effects, whereas the peculiar nature
of spin 3/2 states usually prevents exchange enhancement of
in-plane HH spin splittings in the 2D case \cite{winkler2005}.
However, both the Zeeman splitting and its exchange enhancement
remain suppressed for directions perpendicular to the wire, due to
the same reasons causing this suppression for in-plane field
directions in a 2D HH system. This is in clear contrast to the
electron systems where exchange enhancement of Zeeman splitting is
isotropic \cite{thomas1996}.

In conclusion, we have studied the Zeeman splitting of hole
quantum wires in the ballistic regime. We uncovered a very strong
anisotropy of the effective Land\'{e} \emph{g}-factor for $B$
parallel and perpendicular to the quantum wire. The 1D confinement
significantly increases the anisotropy existing in the 2D HH
system, known as a consequence of the SO coupling. This result
opens a new way to engineer spin-splitting in 1D nanostructures.

R. D. thanks S. E. Andresen, N. Kemp and D. MacGrouther for useful
comments and proof reading of the paper. U. Z. thanks O. P.
Sushkov for enlightening discussions. U. Z. is partially supported
by the Marsden Fund of the Royal Society of New Zealand and
acknowledges a Gordon Godfrey Fellowship from UNSW. M. Y. S.
acknowledges an Australian Research Council Federation Fellowship.
This work has been funded by the Australian Research Council and
the EPSRC.


\begin{thebibliography}:
\bibitem{zutic2004} I. $\check{\mathrm{Z}}$uti$\acute{\mathrm{c}}$ , J. Fabian and S. Das Sarma ,
Rev. Mod. Phys. {\bf 76}, 323 (2004). 
\bibitem{byshkov1984} Yu. A. Bychkov and E. I. Rashba, J. Phys.
C: Solid State Phys., {\bf 17}, 6039 (1984).
\bibitem{datta1990} S. Datta and B. Das, Appl. Phys. Lett. {\bf 56}(7), 665 (1990).
\bibitem{winkler2003} R. Winkler, \emph{Spin-Orbit Coupling Effects in Two-Dimensional Electron and Hole Systems} (Springer, Berlin, 2003).
\bibitem{vankeresten1990} H. W. van Kesteren, E. C. Cosman, W. A. J. A. van der Poel, and C. T.
Foxon, Phys. Rev. B {\bf 41}, 5283 (1990).
\bibitem{patel1991a} N. K. Patel \emph{et al.}, Phys. Rev. B {\bf 44}, R10973 (1991).
\bibitem{thomas1996} K. J. Thomas \emph{et al.}, , Phys. Rev. Lett. {\bf
77}, 135 (1996). 
\bibitem{thomas1998} K. J. Thomas \emph{et al.}, Phys. Rev.
B {\bf 58}, 4846 (1998).
\bibitem{daneshvar1997} A. J. Daneshvar \emph{et al.}, Phys. Rev. B {\bf 55},
R13409 (1997). 
\bibitem{papadakis2000} S. J. Papadakis, E. P. De Poortere, M. Shayegan, and R. Winkler,
Phys. Rev. Lett. {\bf 84}, 5592 (2000).
\bibitem{winkler2000} R. Winkler, S.
J. Papadakis, E. P. De Poortere, and M. Shayegan,  Phys. Rev.
Lett. {\bf 85}, 4574 (2000).
\bibitem{note0} In 2D HH systems, measurements of $g^{*}$ are not possible due to the complexity of the hole band structure (see ref \cite{winkler2003} p.
145 for more complete explanation): The anisotropic Zeeman effect
between $[\bar{2}33]$ and $[01\bar{1}]$ is observed by measuring
the magnetic field $B^{*}$ when the 2D HH is fully spin polarized.
Experimental results in these systems found
$B^{*}_{[\bar{2}33]}$/$B^{*}_{[01\bar{1}]} \approx 1/2$ and theory
predicted $g^{*}_{[\bar{2}33]}$/$g^{*}_{[01\bar{1}]}$ $\approx 4$
\cite{papadakis2000,winkler2000,winkler2003} (for a 20 nm quantum
well). This anisotropy is independent of the current direction.
\bibitem{danneau2005} R. Danneau \emph{et al.}, Appl. Phys. Lett. {\bf 88}, 012107 (2006).
\bibitem{note1} The 2D HH system is confined in a 20 nm quantum well and has density 1.2 $\times 10^{15} \mathrm{m}^{-2}$ and mobility 92 $\mathrm{m}^{2}\mathrm{V}^{-1}\mathrm{s}^{-1}$.
\bibitem{note2} Note that the conductance anomaly at $0.7 \times 2e^{2}/h$, which was observed in
similar devices \cite{danneau2005}, does not appear for back and
middle gates voltage (2.5 and -0.5 V, respectively) used in the
present experiment. The presence of the conductance anomaly is
dependent to these gates voltage which contribute to modify the
shape of the confinement potential.
\bibitem{fang1968} F. F. Fang, and P. J. Stiles  Phys. Rev.
{\bf 174}, 823 (1968).
\bibitem{patel1991} N. K. Patel \emph{et al.}, Phys.
Rev. B {\bf 44}, 13549 (1991). 
\bibitem{glazman1989} L.I. Glazman and A.V. Khaetskii,
Europhys. Lett. {\bf 9}(3), 263 (1989).
\bibitem{martinmoreno1992} L. Martin-Moreno, J. T. Nicholls, N. K. Patel, and M.
Pepper, J. Phys.: Condens. Matter {\bf 4}, 1323 (1992).
\bibitem{zulicke2006} U. Z\"{u}licke and O. P. Sushkov (in preparation).
\bibitem{winkler2005} R. Winkler \emph{et al.}, Phys. Rev. B {\bf 72}, 195321 (2005).
\bibitem{note3} These data are taken from another cool-down than shown in Fig.\
1. $g^{*}$ is extracted from Zeeman splitting and
$V_{\mathrm{SD}}$ spectroscopy taken from the same cool-down. To
verify that the confining potential remains constant from one cool
down to the next, we measured the 1D subband spacings each time
the device was thermally cycled. These were constant to within our
experimental accuracy $\pm$ 5$\%$.
\end{thebibliography}
\end{document}